\documentclass[iop]{emulateapj}

\usepackage{amssymb}
\usepackage{color}
\usepackage{graphicx}
\begin{document}

\title{Testing Density Wave Theory with Resolved Stellar Populations around Spiral Arms in M81}

\author{Yumi Choi\altaffilmark{1}, Julianne J. Dalcanton\altaffilmark{1}, Benjamin F. Williams\altaffilmark{1}, Daniel R. Weisz\altaffilmark{1,5}, Evan D. Skillman\altaffilmark{2}, Morgan Fouesneau\altaffilmark{3}, and Andrew E. Dolphin\altaffilmark{4}}

\altaffiltext{1}{Department of Astronomy, University of Washington, Box 351580, Seattle, WA 98195, USA; ymchoi@astro.washington.edu}
\altaffiltext{2}{Minnesota Institute for Astrophysics, University of Minnesota, 116 Church Street SE, Minneapolis, MN 55455, USA}
\altaffiltext{3}{Max Planck Institute for Astronomy, Koenigstuhl 17, 69117 Heidelberg, Germany}
\altaffiltext{4}{Raytheon, 1151 E. Hermans Road, Tucson, AZ 85756, USA}
\altaffiltext{5}{Hubble Fellow}
\shorttitle{Testing Density Wave Theory in M81}
\shortauthors{Choi et al.}

\begin{abstract}
Stationary density waves rotating at a constant pattern speed $\Omega_{\rm P}$ would produce age gradients across spiral arms. We test whether such age gradients are present in M81 by deriving the recent star formation histories (SFHs) of 20 regions around one of M81's grand-design spiral arms. For each region, we use resolved stellar populations to determine the SFH by modeling the observed color-magnitude diagram (CMD) constructed from archival {\it Hubble Space Telescope} (HST) F435W and F606W imaging. Although we should be able to detect systematic time delays in our spatially-resolved SFHs, we find no evidence of star formation propagation across the spiral arm. Our data therefore provide no convincing evidence for a stationary density wave with a single pattern speed in M81, and instead favor the scenario of kinematic spiral patterns that are likely driven by tidal interactions with the companion galaxies M82 and NGC 3077. 
\end{abstract}


\section{INTRODUCTION}
\label{sec:introSec}
More than half of the galaxies in the Local Universe show spiral patterns highlighted by young stars  \citep[e.g.,][]{nair10, lintott11}. However, the origin of these patterns remains an open question. Several theories have been suggested to explain the underlying mechanisms connecting spiral patterns and star formation. Two models in particular are the most widely accepted: the global density wave theory for long-lived arms \citep{lindblad60, linshu64} and the local gravitational instability model for short-lived transient arms \citep[e.g.,][]{goldreich65, julian66, sellwood84, sellwood11, elmegreen11}. The density wave theory explains active star formation in the coherent spiral arms as being a result of gas compression by a density wave propagating through a galactic disk. In contrast, the local gravitational instability model predicts that star formation is stochastic, showing no clear age gradients across the spiral arm. However, a wide range of observed appearances of spiral structures, from grand-design to flocculent, implies that a single model may not fully explain the formation and evolution of all type of  spiral patterns. For a detailed review of spiral structures, we refer the reader to \citet{dobbs14} and references therein.

Although multiple mechanisms may be responsible for the variety of spiral patterns observed, it is straightforward to test the spiral density wave theory in individual systems. Specifically, one should be able to find a systematic spatial ordering among SF/gas tracers with different timescales (e.g., H$\,\textsc{i}$ for the cold dense gas, 24~$\mu$m for obscured stars, and H$\alpha$ for the young stars) within spiral arms that are supported by quasi-static density waves \citep{roberts69}. In this model, one assumes that density waves are propagating through a galactic disk with a constant angular pattern speed, and that the corotation radius, $R_{\rm cr}$, is defined to be where the spiral pattern speed becomes the same as the rotational speed of the disk. Inside $R_{\rm cr}$, materials move faster than the density waves, and vice versa outside $R_{\rm cr}$. When atomic or diffuse gas enters into a spiral arm (i.e., falling into a spiral potential well), it experiences a shock. The resulting compression naturally leads to the formation of molecular clouds, enhancing star formation. Newly formed stars disperse the surrounding molecular clouds through stellar feedback and continue to move away from the spiral arm. The resulting signature of this process is a spatial sequence of cold molecular gas, obscured star formation, massive young O/B stars, and evolved stars from the upstream to the downstream inside $R_{\rm cr}$, and the same sequence, but with the opposite angular variation, outside $R_{\rm cr}$. 

With grand-design spiral galaxies, many efforts have been made to measure such angular offsets among different SF/gas tracers. However, findings and conclusions on the angular offsets have conflicted even for the same galaxies (e.g., M51 and M74). While \citet{foyle11} found no evidence for a systematic ordering of different tracers in these galaxies, \citet{tamburro08} and \citet{egusa09} detected the expected systematic angular offsets as a function of galactic radius. However, the amplitudes of the measured angular offsets are different from each other and thus the estimated timescales for the total star formation processes are also different. For example, \citet{tamburro08} and \citet{egusa09} reported different SF timescales of 1--4~Myr and 5--30~Myr based on the angular offset measurements between H$\,\textsc{i}$ and 24~$\mu$m and CO and H$\alpha$, respectively. Among various possible causes for the differences, \citet{louie13} showed that the discrepancies are mainly caused by differences in the choice of SF/gas tracers used to measure the angular offsets. They also concluded that CO emission traces the compressed gas better than H$\,\textsc{i}$ 21~cm emission, since a portion of H$\,\textsc{i}$ emission may come from the photo-dissociated gas rather than the direct precursor of SF. In addition, 24~$\mu$m emission may also be contaminated by the underlying older stellar populations, which are not associated with the recent SF activity \citep[e.g.,][]{murphy11, leroy12}. 

In this work, we test the density wave theory by looking for evidence of star formation propagation across the spiral arm in M81 using CMD-based SFH analysis, rather than using angular offset measurements among different SF/gas tracers. This approach avoids any complications related to the choice of SF/gas tracers or to the details of emission peak finding techniques and angular offset measurements from observations of gas emission. 

We carry out this analysis in M81, which is one of the largest disk galaxies in the Local Volume. It is nearly face-on \citep[$i = $ 59$\arcdeg$;][]{deblok08} and has 2 symmetric grand-design spiral arms with low foreground extinction \citep[A$_{\rm V} =$ 0.266;][]{schlegel98}. Thus, M81 is a good laboratory to test theories of spiral formation in detail. To compare our SFH analysis with predictions from the density wave theory, we adopt a high quality H$\,\textsc{i}$ rotation curve from THINGS \citep{deblok08} and explore a variety of pattern speeds reported in the literature for M81. The average of the $R_{\rm cr}$ values found in the literature is $\sim$11.24~kpc (see Table~\ref{tbl1}), and the corresponding pattern speed is $\sim$19~km~s$^{-1}$kpc$^{-1}$ when combined with the rotation curve. Throughout this paper, we assume a distance of $\sim$3.8~Mpc (i.e., $m$ -- $M =$ 27.9), which was derived based on deep resolved photometry of the outer disk of M81 \citep{williams09}. At this distance, an angular separation of 1$\arcsec$ corresponds to physical distance of $\sim$18.5~pc. 

In Section~\ref{sec:DataSec}, we briefly describe properties of the M81 imaging data used in this study. Section~\ref{sec:SFHSec} gives a description of our methodology for deriving the spatially-resolved SFH of spiral arms. In Section~\ref{sec:AnalysisSec}, we present and discuss our results. We summarize our conclusions in Section~\ref{sec:concSec}.

% -------------- TABLE 1 -----------------
\begin{table}
\begin{center}
\caption{A list of $R_{\rm cr}$ values for M81 found in the literature. Since different measurements were made under different assumptions about distance to M81, we converted all values to those at the distance of 3.8~Mpc. The mean $R_{\rm cr}$ is $\sim$11.24~kpc.}
\label{tbl1}
\begin{tabular}{lc} \hline \hline
Reference & $R_{\rm cr}$ (kpc)\\
\hline
\citet{gottesman75} & 12.26--13.25 \\
\citet{rots75}           & 12 \\
\citet{roberts75}      &  11.5 \\
\citet{visser80}        &  13 \\
\citet{sakhibov87}    &  $>$12.86 \\
\citet{elmegreen89} &  9.84 \\
\citet{lowe94}         &  10.52 \\
\citet{westpfahl98}  &  9.8 \\
\citet{kendall08}      &  12.67 \\
\citet{tamburro08}   &  9.23 \\
\citet{feng14}         &  9.5 \\ 
\hline
\end{tabular}
\end{center}
\end{table}

%------------------ Figure 1 --------------------------
\begin{figure} [tbp]
 \begin{center}
      \includegraphics[trim=1.5cm 1cm 0cm 1.5cm, clip=true, height=9cm]{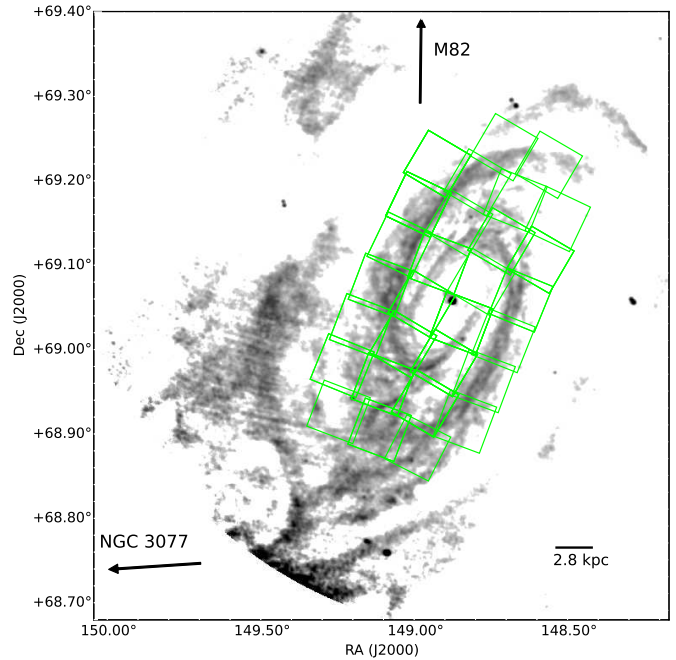}
      \caption{\label{hi_hst} The HST footprints (green boxes) of 29 fields, which cover the entire optical extent of 
      M81, are superposed on the H$\,\textsc{i}$ image \citep[THINGS;][]{walter08}. H$\,\textsc{i}$ emission 
      is extended far beyond optical emission. North is up. East is left. The arrows point out the directions toward 
      companions, M82 and NGC 3077. Tidal bridges connecting M81 and companions are prominent while the 
      tidal features are not seen in the optical.}
   \end{center}
\end{figure}

%------------------ Figure 2 --------------------------
\begin{figure*} [tbp]
 \begin{center}
      \includegraphics[trim=4cm 2.5cm 3cm 0cm, clip=true, angle=270, width=21cm]{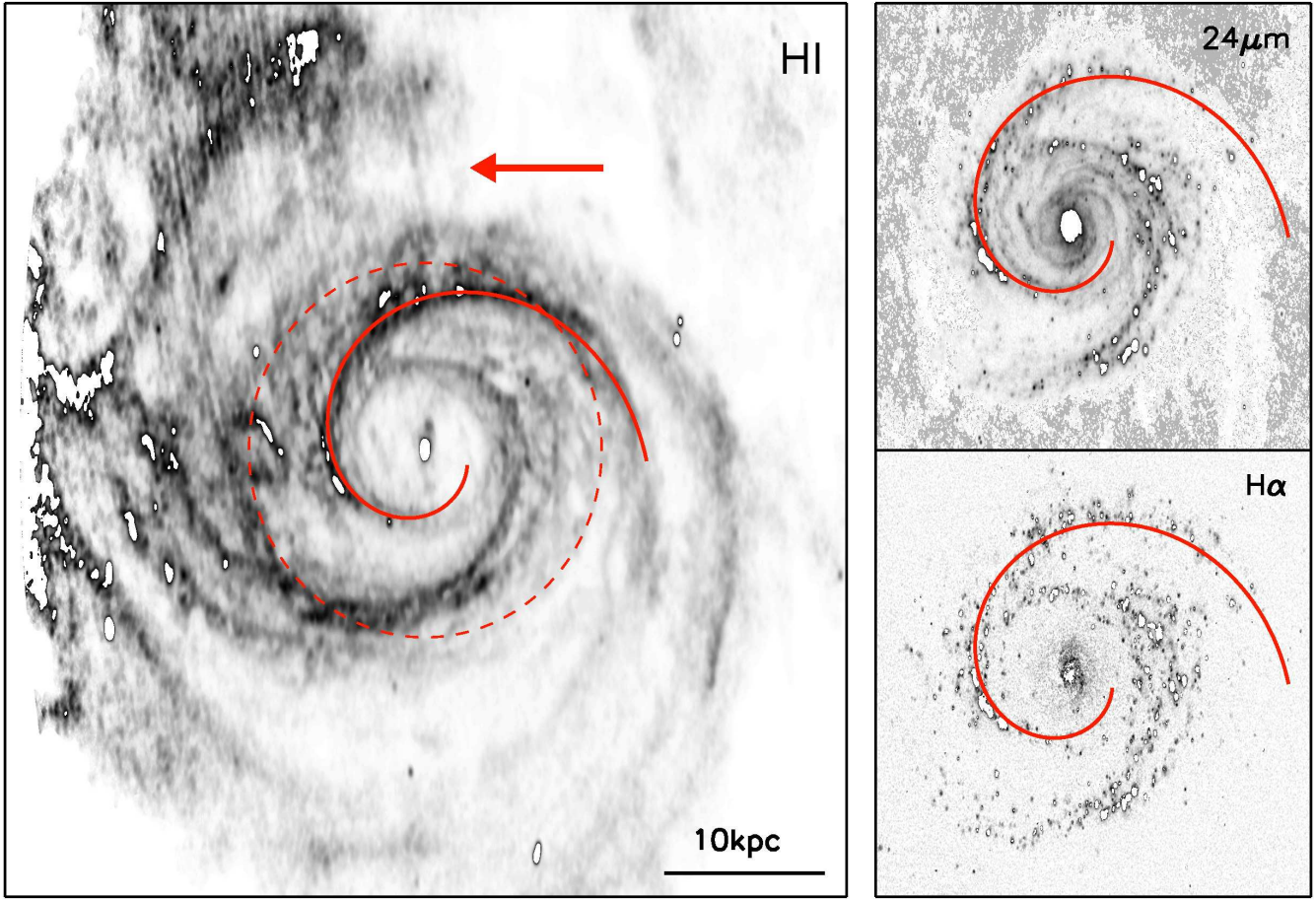}
      \caption{\label{depro} Deprojected image of M81 in H$\,\textsc{i}$ \citep[left, from][]{walter08}, 
      24~$\mu$m \citep[right top, from][]{dale09}, and H$\alpha$ \citep[right bottom, from][]{hoopes01}. 
      An arrow denotes the direction of galactic rotation when assuming a trailing spiral pattern. 
      On each image, we overplot the best-fit logarithmic spiral pattern derived from the H$\,\textsc{i}$ (red solid 
      line) of the NE arm that we analyze in this work. We also mark the reported average 
      corotation radius at $\sim$11.24~kpc on the H$\,\textsc{i}$ image (dashed circle).}
   \end{center}
\end{figure*}

\section{DATA}
\label{sec:DataSec}
We use 29 fields of optical M81 archival data (Proposal ID 10584; Zezas). All fields were imaged in at least two filters (F435W and F606W, including some F814W observations in outer fields). These fields are contiguous and cover the entire optical extent of the galaxy. After making bias and flat-field corrections using the STScI ACS pipeline OPUS, photometry was performed with the ANGST pipeline \citep{dalcanton09}. The pipeline is optimized for stellar photometry on ACS images using the ACS module within the DOLPHOT photometric package \citep{dolphin00}. Briefly, the cosmic rays were identified on a combined single drizzled image using the {\tt multidrizzle} task \citep[PyRAF;][]{koekemoer02} before measuring flux of individual stars. DOLPHOT modifies the Tiny Tim PSF \citep{krist95} to account for the effects on the PSF shape of the telescope temperature changes during orbit. To quantify systematic differences between the model and true PSF, DOLPHOT determines aperture corrections using the most isolated stars in each field. A detailed description of the photometry technique can be found in \citet{dalcanton09}. The final catalog utilized in this study contains objects flagged as stars with S/N $>$ 4 in both F435W and F606W filters that pass the sharpness cuts, ({\tt sharp}$_{\rm F435W}$ + {\tt sharp}$_{\rm F606W}$)$^{2} <$ 0.075, and the crowding cuts, {\tt crowd}$_{\rm F435W}$ + {\tt crowd}$_{\rm F606W} <$ 0.6. These cuts ensure high quality photometry  by selecting point-like sources that are not significantly affected by crowding \citep{gogarten09}. Using DOLPHOT, we also perform artificial star tests to estimate photometric completeness and uncertainties as a function of magnitude. 

We use the H$\,\textsc{i}$ image from THINGS \citep{walter08} to identify the location of M81's gaseous spiral arms. THINGS provides high-quality 21~cm emission line maps with an angular resolution of $\sim $6$\arcsec$ and spectral resolution of 2.6~km~s$^{-1}$. The beam size of the robust weighted H$\,\textsc{i}$ maps is $\sim$7.5$\arcsec$, corresponding to $\sim$140~pc at the distance of 3.8~Mpc. Figure~\ref{hi_hst} shows the H$\,\textsc{i}$ map overlaid with the footprints of 29 HST fields. We also use the H$\alpha$ \citep[KPNO;][]{hoopes01} and 24~$\mu$m \citep[Spitzer;][]{dale09} images to provide a quick comparison between gas/SF tracers in Section~\ref{sec:DefArmSec} (see Figure~\ref{depro}).

\section{THE SPATIALLY RESOLVED STAR FORMATION HISTORY}
\label{sec:SFHSec}
 \subsection{Defining Spiral Arms and Selection of Stars}
\label{sec:DefArmSec} 

\begin{figure} [tbp]
 \begin{center}
      \includegraphics[trim=2.2cm 3.5cm 1.7cm 3.5cm, clip=true, width=9cm]{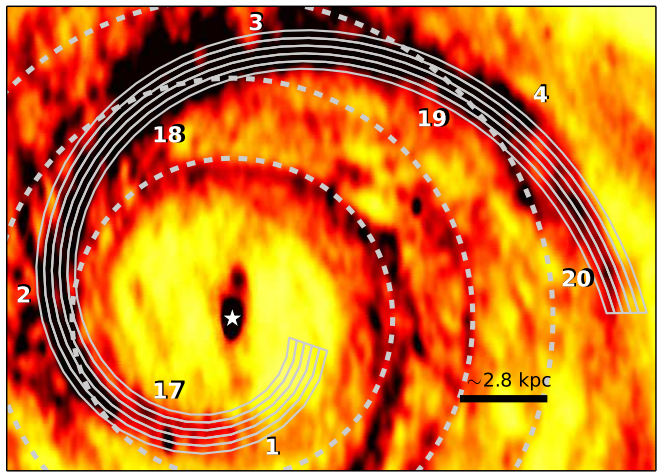}
      \caption{\label{defregions} The 20 analysis regions around the spiral arm are overlaid on the 
      deprojected H$\,\textsc{i}$ map. Individual regions are bounded by dashed circles and solid  
      curves. Dashed circles are located at $\sim$5.6~kpc, $\sim$8.4~kpc, and $\sim$11.2~kpc from the 
      galactic center. To clarify our numbering system, we show nominal numbers for regions located at the 
      most leading and most trailing stripes.}
   \end{center}
\end{figure}

To define the spiral arms, we first deproject the H$\,\textsc{i}$ image assuming a position angle of 330$\arcdeg$ and inclination of 59$\arcdeg$ \citep{deblok08} (left panel in Figure~\ref{depro}) and then select density maxima positions by picking the center of contours within the spiral arms. After translating each ($x$, $y$) position to polar coordinates ($r$, $\theta$), we fit the selected density maxima positions with a logarithmic function of the form ln($r$) $=$ ln($a$) + $b\,\theta$. The azimuthal angle ($\theta$) increases with the direction of galactic rotation, which is counterclockwise for M81 since we assume a trailing spiral pattern. The pattern is almost a perfect logarithmic spiral and the derived pitch angle of the H$\,\textsc{i}$ gas arm, $i = \arctan (|b|)$, is $\sim$15$\arcdeg$, which is in good agreement with previous studies \citep[e.g.,][]{rots75, kennicutt81, lowe94, puerari14}. We adopt this best-fit logarithmic spiral as the peak of the gaseous spiral arms. 

M81 has 2 strong spiral arms, separated by $\sim$180$\arcdeg$. We focus on the NE arm (highlighted with the red solid line in Figure~\ref{depro}), which does not have a strongly streamed feature in the H$\,\textsc{i}$ map  compared to the other stretched arm (SW arm). This choice should minimize the effects of any ongoing tidal interaction on SFH, although the SW arm has a similar pitch angle in the H$\,\textsc{i}$ gas map. The best-fit logarithmic spiral is also overlaid on the deprojected 24~$\mu$m (right top) and H$\alpha$ (right bottom) images. Although the observed peaks of both H$\alpha$ and 24~$\mu$m emission are mostly seen downstream of the best-fit spiral arm, they are scattered and discrete. This observational discreteness can cause complexity and ambiguity in measuring and interpreting the angular offsets among different SF tracers. Note that in both the H$\,\textsc{i}$ and H$\alpha$ maps, the spiral pattern is hardly seen in the inner galactic region ($r \lesssim$ 3~kpc), possibly indicating the inner Lindblad resonance. In contrast, the 24~$\mu$m image shows a signature of short spiral arms in the inner galactic region in addition to the grand-design spiral arms, as first recognized both in the 3.6~$\mu$m and 8~$\mu$m images by \citet{kendall08} and explored in more detail by \citet{feng14}. The driving forces of these short spiral arms is beyond the scope of this paper.

Figure~\ref{defregions} shows our 20 (5 $\times$ 4) analysis regions around the best-fit logarithmic spiral arm. Individual regions are bounded by dashed circles, which denote galactic radii of $\sim$5.6~kpc, $\sim$8.4~kpc, and $\sim$11.2~kpc, and solid curves, which define 5 spiral-shaped stripes. These 20 regions enable us to obtain the spatially resolved SFHs around the spiral arm using sufficient numbers of stars in each region, leading to reliable SFHs. We first define 5 spiral-shaped stripes along the best-fit logarithmic spiral arm based on the perpendicular distance from the best-fit spiral arm (from --700 to $+$700~pc with a constant width of 280~pc). The total width of 1.4~kpc is wide enough to catch the potential age gradients across the spiral arm predicted by the density wave theory (see Section~\ref{sec:SFpropSec} for timescales of interest).  Each spiral-shaped stripe is then further divided into 4 different bins based on the distance to the galactic center (from $\sim$2.8~kpc to $\sim$13.8~kpc with a constant radial length $\Delta\,r\sim$2.8~kpc). The innermost and outermost radial boundaries are chosen to limit the analysis to the regions where the bulge components are negligible and the spiral pattern is clear. By choosing $\Delta\,r\sim$ 2.8~kpc, the boundary between the third and the fourth bins is aligned with the reported average R$_{cr}$ of $\sim$11.24~kpc, and thus the fourth bin is likely outside R$_{cr}$. Therefore, in the case of $\Omega_{\rm P} =$ 19~km~s$^{-1}$kpc$^{-1}$, one can avoid the potential ambiguity arising from the coexistence of two oppositely propagating SF within a single bin. However, for a pattern speed faster (slower) than 19~km~s$^{-1}$kpc$^{-1}$, corotation radius lies within the third (fourth) radial bin. We assign nominal numbers to these 5 $\times$ 4 regions. Smaller numbers are assigned to stripes with larger azimuthal angle (i.e., the leading side within R$_{\rm cr}$, but the trailing side beyond R$_{\rm cr}$). Within each stripe, the numbers increase with increasing galactic radius.  

\begin{figure*} [tbp]
 \begin{center}
 \begin{picture}(0,0)
 \thicklines
 \put(170,490){Increasing galactic radius}
 \put(270,492){\color{red}\vector(1,0){100}}
 \put(455,335){\rotatebox{-90}{Decreasing azimuthal angle}}
 \put(458,230){\color{red}\vector(0,-1){100}} 
 \end{picture}
      \includegraphics[trim= 1cm 0cm 0cm 0.5cm, clip=true, height=17.5cm]{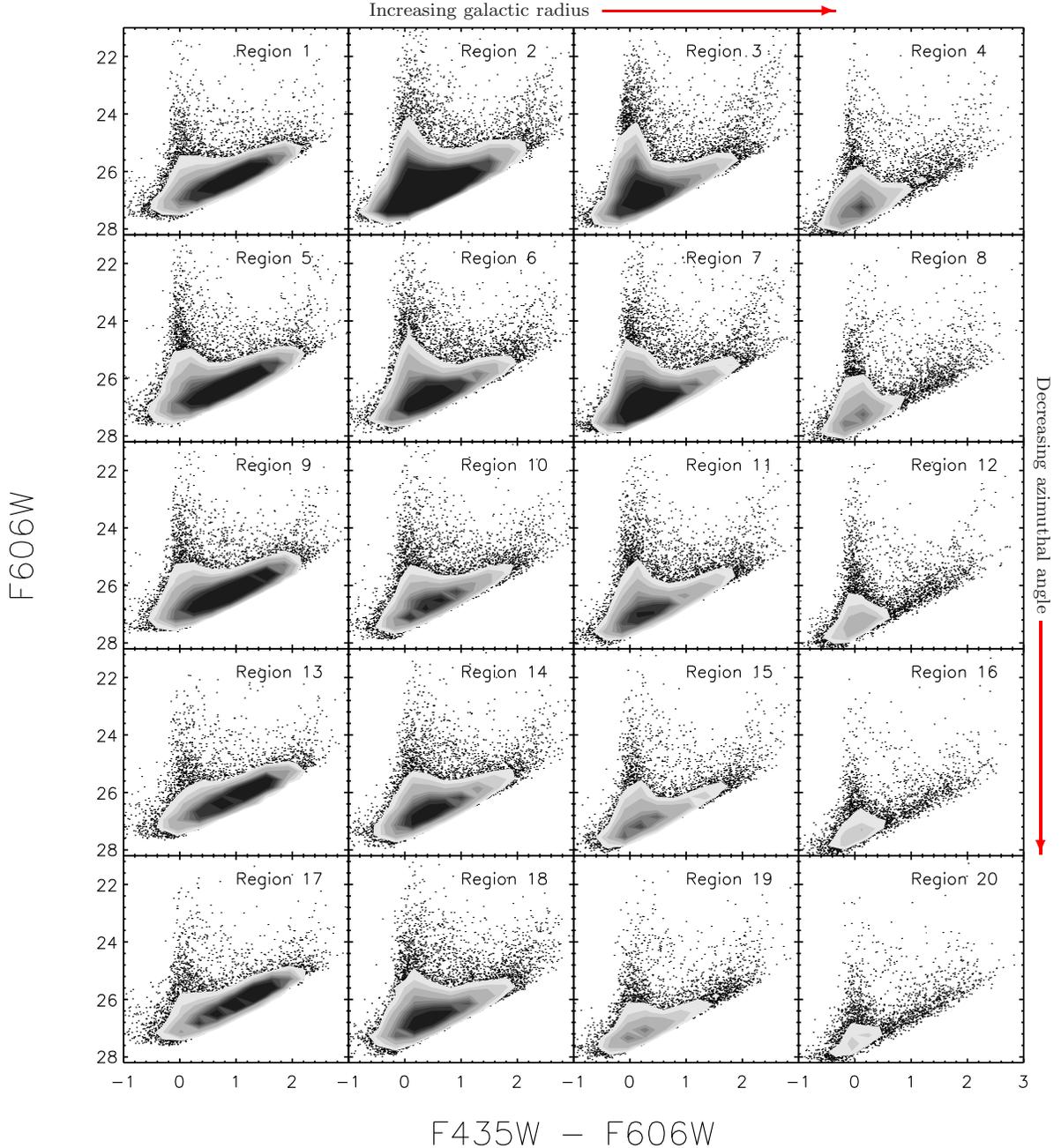}
      \caption{\label{cmd} Observed CMDs of 20 regions. Each column represents the different radial bins in 
	the spiral arm. The galactic radius increases from left to right. Within each radial bin, azimuthal 
	angle decreases from top to bottom panels. All CMDs have significant populations of upper main sequence 
	stars, which are critical for constraining recent SFH.}
   \end{center}
\end{figure*}

The numbers of stars, 50\% photometric completeness limits (derived from artificial star tests), and the areal coverage for individual regions are listed in Table~\ref{tbl2}. All regions have almost the same depth and similar areal coverage (2--3~kpc$^2$). The observed CMDs of each region are shown in Figure~\ref{cmd}. Each column represents different radial bins, with galactic radius increasing from the left to the right columns. Each row represents different spiral-shaped stripes. The azimuthal angle decreases from top to bottom. For example, when adopting $\Omega_{\rm P} =$ 19~km~s$^{-1}$kpc$^{-1}$, in the first 3 columns that are the radial bins likely inside R$_{\rm cr}$, top panels are for the leading side of the arm and bottom panels are for the trailing side of the arm, and vice versa in the last column, which is the radial bin likely outside R$_{\rm cr}$.

% -------------- TABLE 2 -----------------
\begin{table}
\begin{center}
\caption{Numbers of stars, 50\% photometric completeness limits in F435W and F606W bands, the areas covered, and the best-fit $A_{\rm V}$ and $dA_{\rm V}$ values for individual regions. }
\label{tbl2}
\begin{tabular}{ccccccc} \hline \hline
Region & N$_{stars}$ &  F435W  & F606W  & Area  &Av & dAv\\
       &           &  (mag)  & (mag)  & (kpc$^2$) & (mag)  &  (mag)    \\
\hline
1  & 14,850       & 27.07   & 26.93  & 2.93 & 0.30 & 0.9\\
2  & 22,650       & 27.15   & 27.02  & 3.11 & 0.30 & 0.8\\
3  & 18,476       & 27.30   & 27.26  & 3.05 & 0.30 & 0.4\\
4  & 6,578         & 27.49   & 27.48  & 3.41 & 0.25 & 0.1\\
\hline
5  & 16,776       & 27.00   & 26.84  & 3.03 & 0.35 & 0.5\\
6  & 15,978       & 27.20   & 27.07  & 3.01 & 0.40 & 0.8\\
7  & 16,729       & 27.31   & 27.28  & 2.95 & 0.35 & 0.5\\
8  & 6,288         & 27.50   & 27.49  & 3.06 & 0.20 & 0.3\\
\hline
9   & 17,326      & 27.02   & 26.90  & 3.12 & 0.20 & 0.8\\
10 & 12,488      & 27.25   & 27.15  & 2.90 & 0.35 & 1.1\\
11 & 12,438      & 27.37   & 27.34  & 2.84 & 0.30 & 0.6\\
12 & 4,801        & 27.50   & 27.51  & 2.70 & 0.25 & 0.2\\
\hline
13 & 11,942      & 27.06   & 26.89  & 3.17 & 0.30 & 0.9\\
14 & 13,220      & 27.23   & 27.12  & 2.75 & 0.45 & 0.8\\
15 & 7,656        & 27.40   & 27.35  & 2.80 & 0.35 & 0.4\\
16 & 3,350        & 27.55   & 27.57  & 2.36 & 0.20 & 0.3\\
\hline
17 & 12,879      & 27.06   & 26.90  & 3.18 & 0.35 & 0.6\\
18 & 14,890      & 27.21   & 27.13  & 2.65 & 0.35 & 1.0\\
19 & 6,351        & 27.43   & 27.39  & 2.71 & 0.20 & 0.6\\
20 & 2,983        & 27.61   & 27.58  & 2.01 & 0.15 & 0.6\\
\hline
\end{tabular}
\end{center}
\end{table}

\subsection{Deriving the Star Formation Histories}
\label{sec:CMDmodelSec}
In all regions, there are sufficient main sequence stars for age-dating. The SFH of each region is derived using the MATCH package \citep{dolphin02}, which has been applied to many studies \citep[e.g.,][]{williams09, gogarten09, mcquinn10, williams11, weisz11, mcquinn12, weisz14, lewis15}. The fundamental principle behind MATCH is to find a set of synthetic CMDs that best reproduce the observed CMD, using a maximum likelihood method assuming Poisson-distributed data. We use the Padova isochrone set \citep{girardi02} with updated models \citep{marigo08, girardi10} for the CMD fitting. In the actual fitting process, stars brighter than a 50\% completeness limit are used. We run MATCH with a fixed distance of $m$ -- $M =$ 27.9, a Kroupa initial mass function \citep{kroupa01}, and a binary fraction of 0.35, and then solve for the best-fit foreground extinction ($A_{\rm V}$), differential extinction ($dA_{\rm V}$, internal to the galaxy), and linear combination of SFR and [Fe/H] as a function of age. We allow the foreground extinction to range from 0.05 to 0.55~mag with a step size of 0.05~mag and the differential extinction to range from 0 to 1.5~mag with a step size of 0.1~mag. 

We use equally spaced logarithmic time bins between log($\rm t$/yr)~$=$~6.6 and 10.15 with an increment of 0.15~dex. In addition, we require the metallicity to increase monotonically with time, since the observed CMDs are dominated by the upper main sequence which contains little information to constrain metallicity. This metallicity constraint enables us to derive a SFH with a physically plausible metallicity evolution. We estimate the random uncertainties of SFHs by using  the Hybrid Monte Carlo (MC) sampling technique described in \citet{dolphin13}. By creating random samples following the probability density of SFH solutions, the Hybrid MC technique overcomes the issue seen in uncertainties computed from the traditional bootstrap MC resampling, which underestimates the random uncertainties of SFH for time bins with low or zero SFR. The systematic uncertainties, which are dominated by the uncertainty in stellar evolution models \citep{dolphin12}, are only important when comparing the absolute SFH to other galaxies or simulations, but can be safely ignored when comparing regions where CMDs have similar depths within a single galaxy, since all regions will be affected similarly. 

\begin{figure} [tbp]
 \begin{center}
      \includegraphics[trim=1cm 0cm 0cm 1cm, clip=true, width=9.5cm]{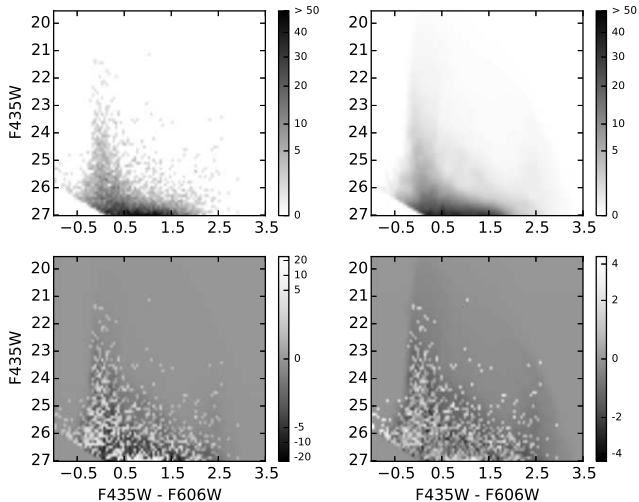}
      \caption{\label{match} Example of the best-fit CMD from MATCH. Upper left: the observed CMD. 
      Upper right: the best-fit model CMD. Bottom left: the residual of data-model CMD. Brighter colors for 
      CMD bins where the model underproduces stars. Bottom right: the residual significance 
      (i.e., variance-weighted residual).}
   \end{center}
\end{figure}

Figure~\ref{match} presents an example of the best-fit CMD from MATCH, along with the observed CMD, the residual after subtracting the model from the data, and the residual weighted by the variance in each CMD bin \citep[residual significance CMD; see][]{dolphin02}. Good agreement between the model CMD and the observed CMD indicates that the derived SFH is acceptable. The best foreground extinction and differential extinction fits are given in the last two columns in Table~\ref{tbl2}. The mean of the best foreground extinction fits is 0.295~mag, and it is consistent with the value of $A_{\rm V} =$ 0.266~mag from \citet{schlegel98}. The mean of the best differential extinction fits is 0.61~mag. The highest differential extinction value ($\sim$1~mag) in each stripe is found in the second radial bin where dust emission is strongest within the arm \citep[Herschel;][]{bendo10}. Without considering differential extinction, we found the best-fit $A_{\rm V} =$ 0.45, which is consistent with the value that was found for the arm region in \citet{williams09}, in which differential extinction was not taken into account. 

\section{Result and Discussion}
\label{sec:AnalysisSec}
\subsection{Recent Star Formation History}
\label{sec:RSFSec}
All regions show similar SFHs at ages $>$ 0.2~Gyr within the uncertainties, as would be expected if old stellar populations have been well mixed azimuthally on timescales longer than the dynamical time and/or mixed radially through radial migration \citep[e.g.,][]{sellwood02, haywood08, roskar08}. Fortunately, the timescales in which we are interested are less than $\sim$0.15~Gyr; M81 has a rotation speed of $\sim$230~km~s$^{-1}$, and a reported pattern speed ranging from $\sim$17--25~km~s$^{-1}$kpc$^{-1}$ \citep{gottesman75, rots75, roberts75, visser80, sakhibov87, elmegreen89, lowe94, westpfahl98, kendall08, tamburro08, feng14}. Given the width of our defined spiral arm, we expect stars to move through our analysis region in less than $\sim$150~Myr at all galactic radii, even where the relative velocity between the density wave and the material is modest ($\sim$30~km~s$^{-1}$). Therefore, we focus only on the recent SFHs.

\begin{figure*}[tbp]
 \begin{center}
 \begin{picture}(0,0)
 \thicklines
 \put(170,490){Increasing galactic radius}
 \put(270,492){\color{red}\vector(1,0){100}}
 \put(455,335){\rotatebox{-90}{Decreasing azimuthal angle}}
 \put(458,230){\color{red}\vector(0,-1){100}}
 \put(365,461){(SFR$\times$5)}
 \put(365,372){(SFR$\times$5)}
 \put(365,283){(SFR$\times$5)}
 \put(365,194){(SFR$\times$5)}
 \put(365,105){(SFR$\times$5)}
 \end{picture}
  	\includegraphics[trim= 1cm 0cm 0cm 0.5cm, clip=true, height=17.5cm]{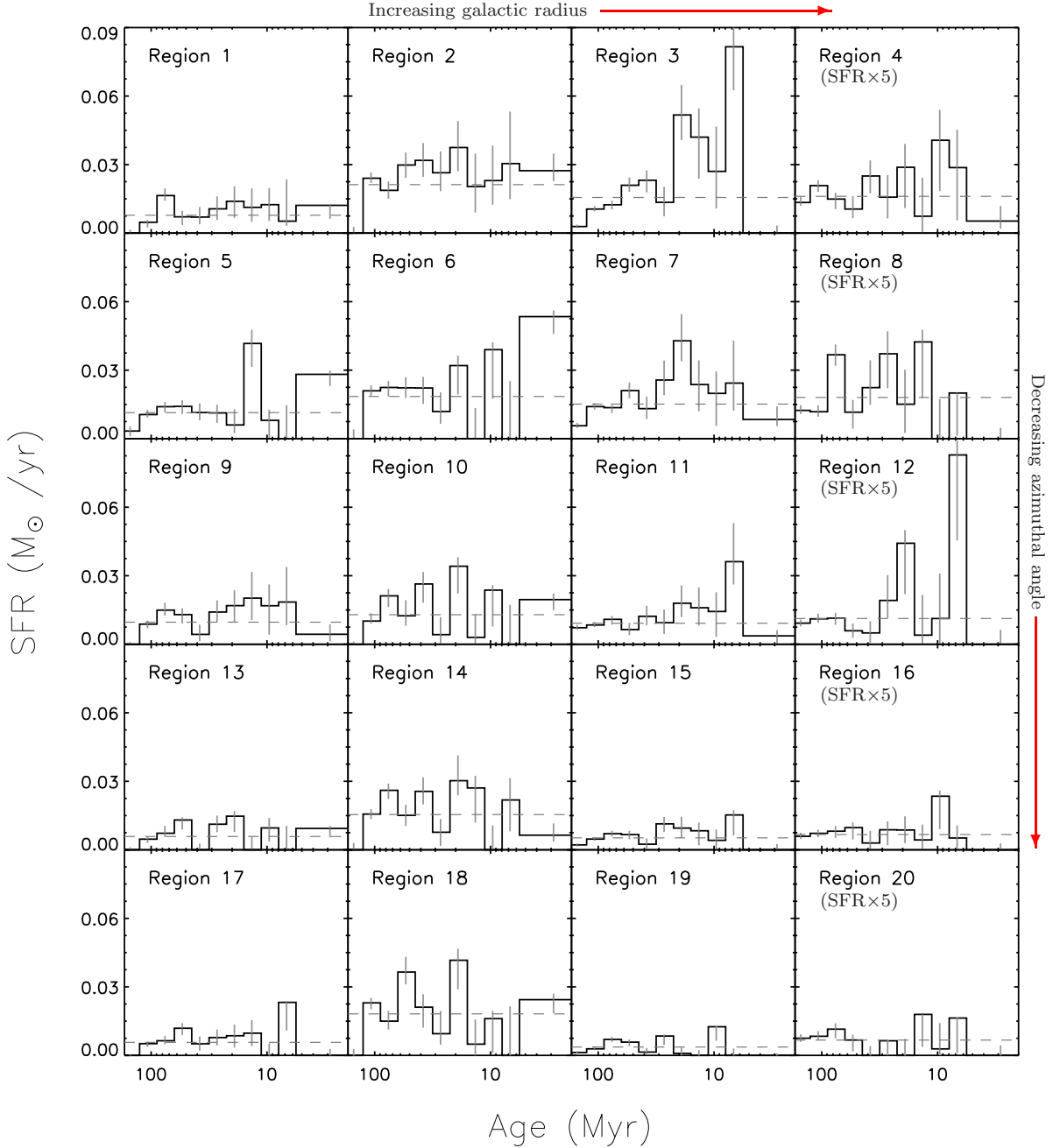}
	\caption{\label{rsfh} The SFHs over the past 150~Myr. Panel order is the same as Figure~\ref{cmd}.
        For clarity, we multiply SFRs of the outermost radial bin (last column panels) by 5. The 
        dashed line marks the average SFR over the past 150~Myr in each region. }
 \end{center}
\end{figure*}

Figure~\ref{rsfh} shows the SFHs of individual regions during the past 150~Myr. In general, the overall SF is weaker at the outer two radial bins, most likely due to the low gas surface density (i.e., the decline in the gas surface density with galactic radius). In the outermost radial bin, a weaker shock might also be responsible for lower SFR around the corotation radius, due to the small relative speed between the gas and the density wave, if it exists. For clarity, we multiply SFRs of the outermost radial bin (last column panels in Figure~\ref{rsfh}) by 5. The dashed line indicates the average SFR over the past 150~Myr in each region. All regions show clear evidence of young stars.

%-------------- Figure 6 ---------------------------------

%-------------- Figure 7 ----------------------------------
\begin{figure} [tbp]
\begin{center}
      \includegraphics[scale=0.5, trim=1cm 13cm 0cm 0.5cm, clip=true]{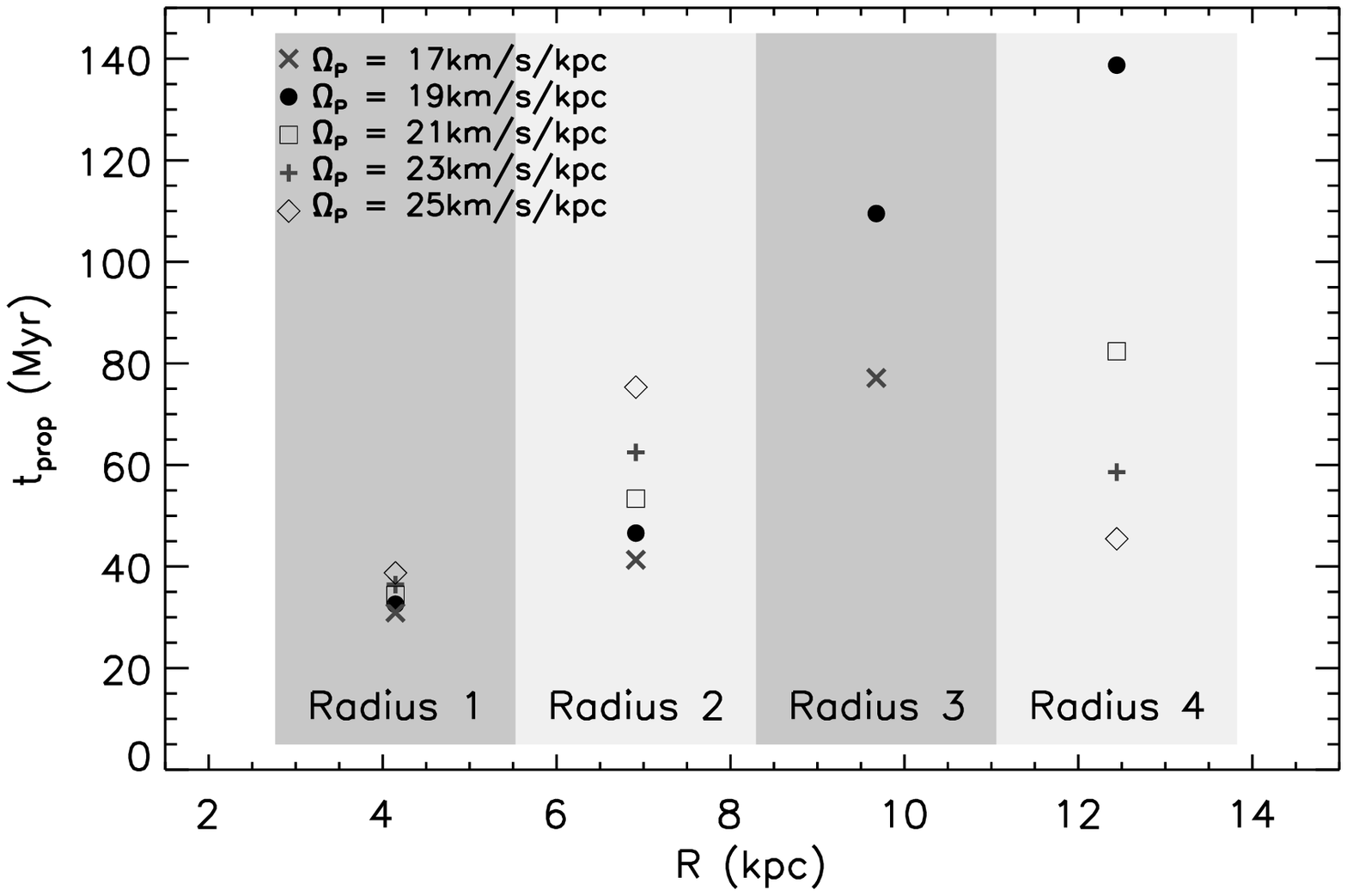}
      \caption{\label{timescale} Predicted timescale of SF propagation across the entire 5 spiral-shaped stripes as a function of galactic radius, for a variety of pattern speeds: 17~km~s$^{-1}$kpc$^{-1}$ (crosses), 19~km~s$^{-1}$kpc$^{-1}$ (filled circles), 21~km~s$^{-1}$kpc$^{-1}$ (open squares), 23~km~s$^{-1}$kpc$^{-1}$ (pluses), and 25~km~s$^{-1}$kpc$^{-1}$ (open diamonds). Shaded boxes denote the extents of the four radial bins. For $\Omega_{\rm\,p} =$ 19~km~s$^{-1}$kpc$^{-1}$, corotation occurs at $\sim$11.24~kpc, which is well aligned with the boundary between the third and fourth radial bins. For $\Omega_{\rm\,p} =$ 17~km~s$^{-1}$kpc$^{-1}$, corotation radius lies within the fourth radial bin while corotation radius for the other three pattern speeds (21, 23, and 25~km~s$^{-1}$kpc$^{-1}$) lies within the third radial bin.}
 \end{center}
\end{figure}

\subsection{Star Formation Propagation}
\label{sec:SFpropSec}
Density wave theory predicts age gradients across spiral arms. \citet{dobbs10b} explored the spatial distribution of young star clusters with different ages ($\sim$2--130~Myr) for four different galaxy models; a galaxy with a stationary density wave, a barred galaxy, a flocculent galaxy, and an interacting galaxy. Their model galaxy with a stationary density wave shows an obvious trend in age of star clusters across its spiral arm, which is a clearly differentiated feature from their other model galaxies. If M81's spiral pattern is truly driven by a traditional density wave, then we expect to find a peak in SF that shifts in age as one moves azimuthally away from the spiral arm. The direction of the shift in age, and its amplitude, depends on the relative angular velocity between the gas and the density wave, which itself depends on galactic radius. 

To explore whether the recent SFHs of M81's spiral arm are consistent with SF propagation predicted by a steady density wave with a single pattern speed, we compare the derived SFHs with the expected age trend across the spiral arm in each radial bin for a given single pattern speed. The expected timescale ($\Delta\,t$) for a given width of azimuthal angle ($\Delta\theta$) is estimated based on a given single pattern speed and the rotation curve amplitude \citep[V$_{\rm C}(r) \approx$ 200--250~km~s$^{-1}$ at the radii we analyze;][]{deblok08}: 
\begin{eqnarray}
\Delta\,t(r) &=& {\frac{\Delta\theta\,(r)}{\Omega\,(r)-\Omega_{\rm\,p}}}\,,
\end{eqnarray}
where $\Omega\,(r) \equiv {\frac{V_{\rm\,C}(r)}{r}}$.

Figure~\ref{timescale} shows the estimated timescale for SF to propagate through the width of each analysis regions as a function of galactic radius for 5 different pattern speeds, $\Omega_{\rm\,p} =$ 17, 19, 21, 23, and 25~km~s$^{-1}$kpc$^{-1}$. In principle, the timescales varies with galactic radius such that it increases toward the R$_{\rm cr}$, becomes infinity at the R$_{\rm cr}$, and decreases beyond the R$_{\rm cr}$. The direction of the SF propagation outside R$_{\rm cr}$ is expected to be opposite to that inside R$_{\rm cr}$. In the case of $\Omega_{\rm\,p} =$ 19~km~s$^{-1}$kpc$^{-1}$, R$_{\rm cr}$ is well aligned with the boundary between our third and fourth radial bins, and thus we are able to calculate the average timescales to cross our defined spiral arm in all radial bins, which are about 32~Myr, 47~Myr, 109~Myr, and 137~Myr for each radial bin (from inner to outer). These timescales are long enough for SF to emerge from dense molecular clouds, which is believed to occur on timescales of $\sim$5--30~Myr \citep[e.g.,][]{elmegreen00, ostriker01, egusa09}. However, as we mentioned above, for a pattern speed faster (slower) than 19~km~s$^{-1}$kpc$^{-1}$, R$_{\rm cr}$ lies within the third (fourth) radial bin. When R$_{\rm cr}$ falls within one of our radial bins, we expect two waves of SF propagating in opposite senses in that single radial bin. Thus, we do not provide the expected timescales for that radial bin (third bin for $\Omega_{\rm\,p} =$ 21, 23, 25~km~s$^{-1}$kpc$^{-1}$, and fourth bin for $\Omega_{\rm\,p} =$ 17~km~s$^{-1}$kpc$^{-1}$) due to the ambiguity mentioned in Section~\ref{sec:DefArmSec}.

\begin{figure*}[tbp]
 \begin{center}
  	\includegraphics[trim= 1.5cm 1.7cm 2cm 2.2cm, clip=true, width=16cm]{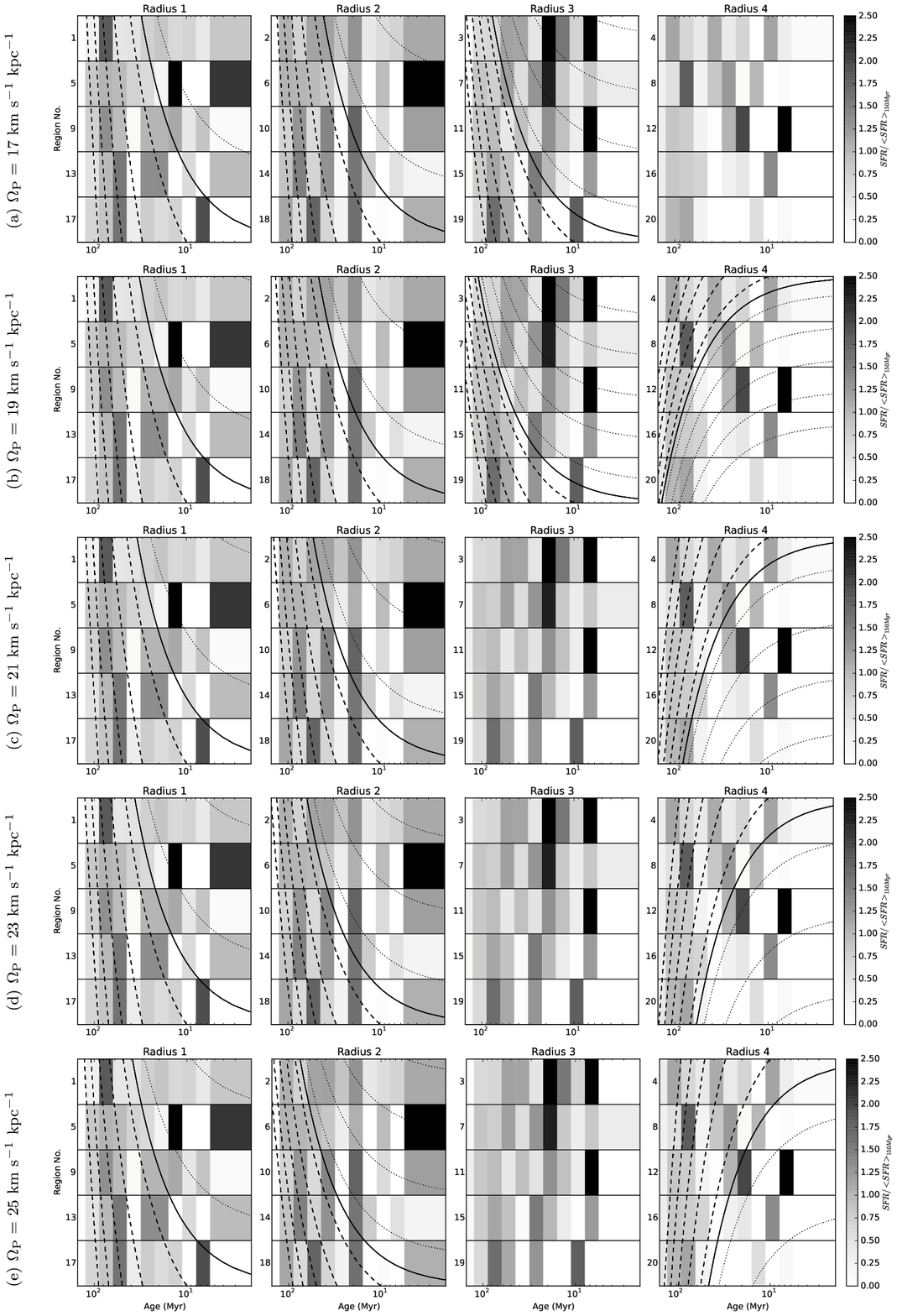}
	\caption{\label{comp} Comparison between our SFHs and the SF propagation predicted 
	by the steady density wave theory. The comparison was carried out in individual radial bins (increasing in 
	galactic radius from left to right) for 5 different pattern speeds (shown in each row): 
	(a) $\Omega_{\rm\,p} =$ 17~km~s$^{-1}$kpc$^{-1}$, 
	(b) $\Omega_{\rm\,p} =$ 19~km~s$^{-1}$kpc$^{-1}$, 
	(c) $\Omega_{\rm\,p} =$ 21~km~s$^{-1}$kpc$^{-1}$, 
	(d) $\Omega_{\rm\,p} =$ 23~km~s$^{-1}$kpc$^{-1}$, and 
	(e) $\Omega_{\rm\,p} =$ 25~km~s$^{-1}$kpc$^{-1}$. 
	Within each panel, each row shows the SFH in a given spiral-shaped stripe; the number of the stripe 
	refers to regions defined in Figure~\ref{defregions}. The SFHs of individual regions are normalized to 
	their average SFR over the past 150~Myr. Grayscale shows the enhanced (darker) or depressed (lighter) 
	SF events compared to the average SFR in each region. The solid curve traces the expected dominant 
	stellar ages across the spiral arm by assuming the onset of SF at the edge of our trailing side spiral-shaped 
	stripe. The dashed (dotted) curves consider positive (negative) time shifts of 10, 30, 50, 70, and 90~Myr. 
	We do not provide the predicted SF propagation in radial bins that are located at corotation, where no 
	dominant propagation is expected. In these plots, a propagating spiral density wave would show enhanced 
	SF that tracks one of the curves in all four radial bins of a single row corresponding to the true pattern 
	speed. We see no evidence for this expected behavior.}
 \end{center}
\end{figure*}

Because of the uncertainty in the pattern speed and the fact that we do not know a priori in which stripe the SFR is expected to peak, we need to test for consistency with a variety of pattern speeds and possible time shifts in the onset of SF. Although our defined 20 regions cover the high-density region of the H$\,\textsc{i}$ gaseous spiral arm, the outer edge of our trailing side stripe does not necessarily mark the spot where SF is initiated by a shock. 

In Figure~\ref{comp}, we therefore compare our derived SFHs with the predicted SF propagation for 5 different pattern speeds and at least 6 different time shifts in the onset of SF to find any systematic trend in the enhanced SF peaks across the spiral arm. Each panel presents the SFHs of 5 adjacent spiral-shaped stripes within a single radial bin, with  the azimuthal angle decreasing from top to bottom (i.e., from leading side to trailing side inside R$_{\rm cr}$, and vice versa outside R$_{\rm cr}$); each grayscale panel therefore corresponds to an entire column of Figure~\ref{rsfh}. Region numbers are given in the y-axis for clarity. The SFH of each region is normalized to its average SFR over the past 150~Myr to look for correlations between SFR enhancements and viable propagating spiral pattern models, and is presented in grayscale where darker (lighter) colors indicate higher (lower) SFR than the average SFR.

In each panel, a solid curve traces the ages of the stellar populations that are predicted to have formed right at the edge of our trailing side spiral-shaped stripe (i.e., a zero time shift in the onset of SF) and moved across the spiral arm if there were a stationary density wave with a given single pattern speed. Dashed (dotted) lines denote the positive (negative) time shifts of 10, 30, 50, 70, and 90~Myr, meaning SF occurred before (after) entering our analysis region. 

Comparing the grayscale to the model lines, the enhanced SF peaks across the spiral arm show no clear systematic trend in all radii to support a steady density wave with a single pattern speed. Instead, the recent SFHs are fluctuating and stochastic. Our finding is robust against both adopting a variety of pattern speeds (i.e., the different SF propagation timescales found in each row of panels) and applying ranges of time shift in the onset of SF (i.e., the dashed and dotted lines in each panel). For the innermost radial bin, there is a weak trend that fits with a time shift of 50~Myr for any given pattern speed cases. However, a SFH with much higher time resolution would be required to determine which pattern speed fits the best with which time shift; our SFH time resolution, especially in older age bins, is not sufficient enough to constrain this. In addition, the amplitudes of some enhanced SF peaks associated with the trend are not strong. 

A successful propagation model must also work at all radii for a single pattern speed. However, there are no clear trends in the outer radial bins that could be explained by any of these curves with the time shift of 50~Myr that possibly fits the innermost radial bin. For example, in the third radial bin, there is one possible trend that shows better agreement either with a zero time shift for $\Omega_{\rm\,p} =$ 17~km~s$^{-1}$kpc$^{-1}$ or with $\Omega_{\rm\,p} =$ 19~km~s$^{-1}$kpc$^{-1}$ with a time shift of --10~Myr. However, these trends are too weak to support any specific pattern speeds and value of time shift, and are not reflected at other radii. Thus, we detect no evidence for the age gradients that would have been expected for SF propagation by a traditional spiral density wave.   

Our results agree with \citet{foyle11}, who measured angular offsets between H$\,\textsc{i}$ and 24~$\mu$m in M81's spiral arm, and found no systematic ordering as a function of galactic radius. Our results also agree with other recent studies that found no significant observational evidence favoring the density wave theory in other nearby grand-design spiral galaxies. For example, \citet{ferreras12} found no offsets between H$\alpha$ and NUV-optical color in NGC 4321. \citet{martinez13} also failed to detect color gradients in the grand-design two-armed spiral galaxies NGC 578, NGC 1703, NGC 4603, NGC 4939, NGC 6907, and NGC 6951.

\subsection{Tidally induced grand-design spiral arm?}
\label{sec:TidalSec}
The absence of a systematic trend in our recent SFHs indicates that the grand-design spiral structure of M81 is not the result of the gas response to long-lived density waves with a single pattern speed. In fact, stationary spiral waves have never been reproduced without imposing very specific conditions in numerical simulations, and spiral patterns are found to be transient/recurrent features \citep[e.g.,][]{sellwood11, fujii11, wada11, grand12,  baba13, sellwood14}. \citet{donghia13} recently showed that non-linear growth of self-induced spiral structures by swing amplifiication can demonstrate `apparent' long-lived spiral arms in their simulation of an isolated galaxy. They suggested that `apparent' long-lived global spiral patterns are merely products of connected self-perpetuating local segments, which are in local dynamical balance between shear and self-gravity. However, the self-perpetuating spiral arms that are locally fluctuating with time do not necessarily exhibit the systematic radial dependence of the relative amplitude of spiral arms that were found in the observed mass surface density map of M81 \citep[e.g.,][]{elmegreen89, kendall08}. 

A snapshot of the current bisymmetric spiral patterns of M81 might be a result of the superposition of two or more $m =$ 2 modes with distinct pattern speeds \citep{sellwood14}. If M81's grand-design spiral arms consist of the superposition of two or more modes lasting 5--10 galactic rotations, then it is reasonable to expect two or more distinct SF propagations across the spiral arm. However, we find no such features in Figure~\ref{comp}. This does not mean that we can decisively rule out this mechanism, since there are possibilities that the relevant pattern speeds are out of our search range or that the superposition of multiple modes would partially erase/overwrite the each other's age gradients, increasing complexity in the resulting SFH.

As an alternative model for grand-design two-armed galaxies, \citet{toomre72}, \citet{kormendy79}, \citet{hernquist90}, and \citet{bottema03} argue that all non-barred grand-design galaxies with typical rotation curves (i.e., flat rotation curves) are likely to be the result of interactions with nearby companions since they frequently reside in the interacting systems showing tidal features such as bridges and tails. These conditions fit M81; it has no strong bar, but has companion galaxies.

Such an interaction is clearly taking place in M81, making tidal interactions a compelling mechanism for driving its spiral arms. Numerical simulations \citep[e.g.,][]{thomasson93, yun99} suggested tidal interactions to explain the current H$\,\textsc{i}$ gas distribution in the M81 group \citep{appleton81}. According to Yun's simulation \citep{yun99}, M81 underwent the closest passage about 220~Myr and 280~Myr ago by its companion galaxies M82 and NGC 3077, respectively. In addition, many observational studies \citep[e.g.,][]{yun94, chandar01, perez06, demello08, konstantopoulos09, santiago10, hoversten11, lim13} have found evidence for tidally induced SF in both M81 and its companion galaxies: ages of compact star clusters and star-forming regions range $\sim$100--500~Myr. Although we also detected enhanced SF at ages between 200 to 350~Myr in all 20 regions, the associated systematic errors are significant since the stars sensitive to this SF peak are near the 50\% completeness limit of the data, and therefore this result would require additional observations to confirm with confidence.

Recent studies \citep[e.g.,][]{oh08, dobbs10a, struck11, oh15} explored the physical properties of simulated galaxies having tidally induced symmetric two-armed grand-design spiral arms. These tidally excited grand-design spirals are almost logarithmic and typically kinematic density waves, which are the result of nested elliptical stellar orbits induced by tidal perturbation. These kinematic density waves do show radial variation of their pattern speed. However, slower differential rotation than the material disk enables the kinematic spirals to survive for a timescale of $\sim$1~Gyr \citep{oh08}. \citet{speights12} estimated the pattern speed of M81's spiral structure and indeed detected its radial dependence, indicating the spiral pattern does actually follow differential rotation.  

Although \citet{dobbs10a} modeled M51 with an interacting companion (NGC 5195), there are some similarities that we can apply to the M81 group as well. During detailed comparison of their model galaxy to HST observations of M51, they noticed that two passages by the companion galaxy at different times are able to  reproduce observed kink/bifurcations in the spiral arm besides a well-defined two-armed spiral pattern. A kink/bifurcation is formed where the younger arm, induced by the second passage, and the older arm, induced by the first passage, are connected. Since M81 is also believed to have experienced two tidal interactions separated by $\sim$60~Myr with two companion galaxies, M82 and NGC 3077, one would expect to observe a similar feature in M81. In fact, bifurcation is seen in the deprojected H$\,\textsc{i}$ image (Figure~\ref{depro}) at around the corotation radius, but the gas density of the bifurcated arm at outer radius is not as high as that of the main NE arm that we have analyzed. The SW arm shows more complicated features, such as at least two bifurcations. 
The existence of the bifurcation indicates the main arm by the earlier passage should be prominent by the time of the later passage. Therefore, if the bifurcation in M81 is a real product of the two separate passages, we are able to infer that the spiral structure of M81 has survived for at least $\sim$280~Myr. \citet{williams09} conducted a SFH analysis for their M81-deep ($\sim$2~mag deeper than our data) field in the outer disk, and also suggested that the M81's spiral arms are at least 100~Myr old.

\section{CONCLUSIONS}
\label{sec:concSec}
In this paper, we investigated the validity of the stationary density wave theory to explain the grand-design spiral structure in M81. We analyzed resolved stellar populations in 20 regions around the logarithmic spiral arm, and derived star formation histories of individual regions using CMD-fitting. Our approach avoids many uncertainties inherent in using discrete SF/gas tracers to measure the angular offsets across the spiral arms. 

To test the assumption that the spiral structure of M81 is driven by a stationary density wave with a single pattern speed, we estimated the timescales for the disk material to cross the spiral arm as a function of galactic radius based on the relative velocity between the rotation curve and a given single pattern speed ranging from 17--25~km~s$^{-1}$kpc$^{-1}$. We then compared the predicted SF propagation with the potential age gradient imprinted in our derived SFHs across the spiral arm. The resulting SFHs shows no systematic age gradient across the spiral arm at all radii. Rather the SFHs are stochastic over the timescales of interest. This result provides convincing evidence that the grand-design spiral structure of M81 is not supported by the stationary density wave with a single pattern speed, but instead is likely supported by a tidally induced kinematic waves due to the interactions with companion galaxies. In addition, we also discuss the bifurcation as additional evidence supporting  the tidally induced kinematic density waves in M81, which will wind up slowly and eventually decay. In conclusion, our recent SFH results agree with other studies that have found observational evidence against the traditional density wave theory in other nearby grand-design spiral galaxies \citep[e.g.,][]{foyle11, ferreras12, martinez13}.

\acknowledgements
We are grateful to the referee for constructive suggestions and comments.
This work was supported by the Space Telescope Science Institute through GO-10945 and GO-11307. Support for DRW is provided by NASA through Hubble Fellowship grants HST-HF-51331.0 awarded by the Space Telescope Science Institute. This research has made use of the NASA/IPAC Extragalactic Database (NED) which is operated by the Jet Propulsion Laboratory, California Institute of Technology, under contract with the National Aeronautics and Space Administration.

% -------------- BEGIN BIBLIOGRAPHY -----------------

\bibliographystyle{apj3}
\bibliography{apj-jour,bibliography}

\clearpage

\end{document}